# GAUGE FIXING ON THE LATTICE AND THE GIBBS PHENOMENON[†]

JEFFREY E. MANDULA

U. S. Department of Energy, Washington, DC  20585,  USA
E-mail: mandula@hep2.er.doe.gov

We discuss global gauge fixing on the lattice, specifically to the lattice Landau gauge, with the goal of understanding the question of why the process becomes extremely slow for large lattices. We construct an artificial "gauge-fixing" problem which has the essential features encountered in actuality. In the limit in which the size of the system to be gauge fixed becomes infinite, the problem becomes equivalent to finding a series expansion in functions which are related to the Jacobi polynomials. The series converges slowly, as expected. It also converges non-uniformly, which is an observed characteristic of gauge fixing. In the limiting example, the non-uniformity arises through the Gibbs phenomenon.

## 1  Introduction

In very broad terms, there are two classes of applications of lattice gauge theory. One is the extraction of the physical predictions of QCD at low energies, where perturbation theory is not trustworthy. The other is learning about the mathematical structure of quantum field theory, especially the Green's functions of QCD, again outside the region in which we can safely trust perturbative analysis. For many applications of the first type, most notably the calculation of the hadronic spectrum, there is no need to specify a gauge. In fact, the original proposal of Wilson was to calculate the Feynman path integral by explicitly summing over all gauges. For problems of the second type, by contrast, the specification of a gauge is intrinsic. Also, the most straightforward way to evaluate the non-perturbative renormalization constants of operators is to work in a fixed gauge. These renormalization constants are physically relevant. They are needed to calculate the absolute scale of the strong interaction matrix elements that multiply the fundamental Standard Model couplings in weak decays and in $K^0 \bar{K}^0$ and $B^0 \bar{B}^0$ mixing.

It is well known that gauge fixing is afflicted with a host of problems. Gribov copies[1] — both the analogues of continuum copies and new lattice artifacts — occur on the lattice. We continue to lack a good understanding of how to treat them in principle, and we also have no general understanding of their practical significance, although in some examples the manner in which they are treated affects the results of calculations.[2] Presumably performing the path integral for a gauge theory within a "Fundamental Modular Domain" is an appropriate method, but we have no operational procedure for finding such a domain.[3]

Gauge fixing to the Landau gauge is also a notoriously slow process, especially for large lattices. While there are a number of techniques that work adequately well on small or moderate sized lattices, for the size lattices currently in common use, all algorithms are unfortunately very inefficient.[4] The goal of the work described here is to get some insight into this algorithmic inefficiency.

## 2  The Lattice Landau Gauge

On the lattice, a general gauge transformation has the structure

$$U_\mu(x) \rightarrow U_\mu^{(G)}(x) \equiv G(x) U_\mu(x) G(x + \hat{\mu})^\dagger \qquad (1)$$

Any gauge condition $f(U) = 0$ is implemented following the Fade'ev-Popov procedure:

$$\begin{aligned}
&\int DU\, e^{-S}\, O(U) \Big|_{f(U) = 0} \\
&= \int DU \int DG\, e^{-S}\, O(U)\, \Delta_{FP}(U)\, \delta(f(U^{(G)})) \quad (2) \\
&= \int DU\, e^{-S}\, O(U^{(G)})
\end{aligned}$$

where here $U^{(G)}$ denotes the gauge transform of $U$ to the gauge $f(U^{(G)}) = 0$.

---

Note that on the lattice the path integral measure remains the same as without gauge fixing. The effect of the Fade'ev-Popov determinant is incorporated not by keeping only configurations of links satisfying the $f(U) = 0$ gauge condition and modifying their relative weights, but by keeping all configurations weighted only by the action, and transforming each to the $f(U) = 0$ gauge.

For the lattice Landau gauge, one formulates the gauge condition as a maximization so as to avoid some lattice artifacts

$$\underset{G}{Max} \sum_{x\mu} Re\, Tr\, U_\mu^{(G)}(x) \qquad (3)$$

At each site, a gauge transformation affects 8 links, so the maximization condition at site $x$ is

$$\underset{G(x)}{Max}\, Re\, Tr\, G(x) \sum_\mu (U_\mu(x) + U_\mu(x - \hat{\mu})^\dagger) \qquad (4)$$

If $G(x) = 1$ satisfies the maximization condition, then we recover the lattice form of the differential condition

$$\Delta_\mu A_\mu(x) \equiv \sum_\mu (A_\mu(x) - A_\mu(x - \hat{\mu})) = 0 \qquad (5)$$

where the gauge potential is

$$A_\mu(x) = \frac{U_\mu(x) - U_\mu^\dagger(x)}{2iag} - Tr\frac{U_\mu(x) - U_\mu^\dagger(x)}{6iag} \qquad (6)$$

The most naive implementation of the Landau gauge condition is to cycle through the lattice, imposing the maximization requirement one site at a time, and repeating the process until the configuration has relaxed sufficiently into a global maximum.[5] Two classes of improvements on this method are widely used. One is overrelaxation[6], that is the replacement $G_{max}(x) \rightarrow G_{max}^w(x)$, $(1 \leq w \leq 2)$, either exactly or stochastically. Another is Fourier preconditioning[7].

The conventional wisdom about the inefficiency of gauge fixing is that the lattice configuration develops "hot spots", that is exceptional poorly fixed points which move from site to site under the effect of repeated gauge fixing sweeps, but which persistent tenaciously. Another part of the conventional wisdom, not actually compatible with the idea of "hot spots", is that it is the longest wavelength modes that relax most slowly.

If one looks in detail at the distribution of the values of the deviation from Landau gauge at each site, using $Tr(\Delta_\mu A_\mu)^2$ as a measure of the distance from perfect fixing, one sees not just a few very poorly fixed sites, but a broad range of deviations with no notable gaps. If one looks in Fourier space, one finds that there is a broad range of relaxation times, with many modes decaying slowly, including, but not limited to, the longest wavelengths.

## 3 Gauge Fixing Convergence

If the link variables $U_\mu(x)$ are close to satisfying the Landau gauge condition, we may expand the gauge transformation needed to satisfy it exactly $G(x) = \exp(i\,g(x))$ to second order, and express the Landau gauge condition as the maximization of a quadratic form

$$\underset{g(x)}{Max}\, Re\, Tr\, i\,[g(x)\, U_\mu(x) - U_\mu(x - \hat{\mu})\, g(x)]$$

$$-\frac{1}{2} Re\, Tr\, [g^2(x)\, U_\mu(x) - U_\mu(x - \hat{\mu})\, g^2(x)] \qquad (7)$$

$$+ Re\, Tr\, [g(x)\, U_\mu(x)\, g(x + \hat{\mu})]$$

Setting the variation of the quadratic form with respect to the generator of gauge transformations $g(x)$ to zero gives a matrix inversion problem with a nearest neighbor structure. The eigenvalues of this matrix control the convergence of relaxation methods, and a two dimensional example illustrates the typical character of the eigenvalues.

The eigenvalues of the tridiagonal $N \times N$ matrix

$$\begin{pmatrix} 1 & 1 & 0 & \cdot & \cdot & \cdot & \cdot \\ 1 & 2 & 1 & \cdot & \cdot & \cdot & \cdot \\ 0 & 1 & 2 & \cdot & \cdot & \cdot & \cdot \\ \cdot & \cdot & \cdot & \cdot & \cdot & \cdot & \cdot \\ \cdot & \cdot & \cdot & \cdot & 2 & 1 & \cdot \\ \cdot & \cdot & \cdot & \cdot & 1 & 2 & 1 \\ \cdot & \cdot & \cdot & \cdot & \cdot & 1 & 1 \end{pmatrix} \qquad (8)$$

are

$$\lambda_m = 2\left(\cos\frac{m\pi}{N} - 1\right) \qquad m = 0, 1, \ldots, N-1 \quad (9)$$

The important points to note are that there is always a zero eigenvalue, and that as $N$ becomes large, many small eigenvalues develop:

$$\lambda_m \cong -\frac{m^2 \pi^2}{2N^2} \qquad m \ll N \quad (10)$$

It is not the single vanishing eigenvalue which is troublesome — its eigenvector could always be projected out — but the multiplicity of small ones.

## 4  Conjugate Gradient Gauge Fixing

Landau gauge fixing is a maximization problem, and there are many methods for addressing such problems. Rather than simply sweeping through the lattice, we may elect to use a global method, which acts on all sites at once. Conjugate gradient is one such. It is a well known, efficient method for minimizing (or maximizing) functions of many variables. In lattice work it is the method of choice for inverting the Dirac operator.

Conjugate gradient is a recipe for finding a sequence of mutually conjugate vectors, that is vectors satisfying

$$(h_i, M h_j) = 0 \qquad (i \neq j) \quad (11)$$

where $M$ is the quadratic form to be extremized. Each vector in the sequence constructed from its predecessors

$$h_{i+1} = M h_i + \sum_{j=0}^{j=i} c_j h_j \quad (12)$$

The initial vector is taken to be in direction of steepest descent from the starting point.

The power of conjugate gradient lies in the fact that the values of calculated coefficients do not change as the number of terms in successive approximations is increased. For minimization problems which are exactly, and not just approximately quadratic forms, conjugate gradient converges exactly in a finite number of steps.

A tractable problem which nonetheless retains the characteristic features of Landau gauge fixing is the minimization of an $N$-dimensional quadratic form with eigenvalues evenly distributed between 0 and 1, starting from a point whose displacement from the exact minimum has an equal projection on each eigenvector of the quadratic form. We know the exact solution, of course, but is it quite instructive to study the convergence of the conjugate gradient algorithm to the exact solution.

The clearest basis in which to analyze the problem is that in which the quadratic form to be minimized is diagonal. This is just a choice of basis, and neither accelerates nor retards the convergence.

$$\text{Min} \sum_{i,j=1}^{N} v_j M_{ij} v_j \qquad M_{ij} = \lambda_i \delta_{ij} = \frac{i}{N} \delta_{ij} \quad (13)$$

In this basis the sequence of approximations is

$$v_i^{[n]} = \sum_{m=0}^{n} c_m h_i^{[m]}$$
$$h_i^{[m+1]} = M_{ij} h_j^{[m]} + \sum_{m'=1}^{m} d_{m'} h_i^{[m']} \qquad h^{[0]} = \begin{pmatrix} 1 \\ 1 \\ 1 \\ \cdot \\ \cdot \\ \cdot \\ \cdot \\ 1 \end{pmatrix} \quad (14)$$
$$c_0 = 1$$

## 5  The $N \to \infty$ Limit

We are concerned with the convergence of the conjugate gradient method as the dimension of the quadratic form becomes very large. Conveniently, in the limit where the dimension goes to infinity we can transform the problem of minimizing the quadratic form into a problem of expanding a function in a series of polynomials. This is accomplished by rescaling the index labeling directions into a continuous variable on the interval $[0,1]$ and replacing sums over indices by finite integrals. The mutually conjugate vectors are replaced by polynomials $P_n(x)$, which are mutually orthogonal with respect to the weight function $w(x) = x$ on this interval. The initial point, which by convention we take as the origin, is the constant function. Explicitly, the mapping is

$$x = i/N$$
$$h_i^{[n]} \to P_n(x)$$
$$\sum_{i=1}^{N} \to \int_0^1 dx$$
$$v_i \to f(x)$$
$$M_{ij} = \frac{i}{N}\delta_{ij} \to x\delta(x-y)$$
$$M_{ij}v_j \to xf(x)$$
$$v_i M_{ij} u_j \to \int_0^1 xf(x)g(x)\,dx \quad v^{[0]} \to f_0 = 1$$
(15)

## 6 The Sturm-Liouville Expansion

*6.1 Statement and Solution*

The convergence of the conjugate gradient minimization becomes the convergence of the sequence of approximations

$$f_n(x) \equiv \sum_{m=1}^{n} c_m P_m(x) \to f(x) = 1 \qquad (16)$$

The polynomials $P_n(x)$ are determined by the two properties that they mutually orthogonal on the interval $[0,1]$ with respect to the weight function $w(x) = x$, and that each successive polynomial is a linear combination of all the foregoing polynomials and $w(x) = x$ times its predecessor.

Such a sequence of functions sounds like the solution of a Sturm-Liouville eigenvalue-eigenvector problem. The equation must have regular singular points at 0 and 1, and a factor of $x$ multiplying the eigenvalue. Such an equation need not have solutions which are polynomials at all. However it is simple to verify that with the following choice of exponents and coefficients:

$$\left[\frac{d}{dx}x^2(1-x)\frac{d}{dx} - 2\right]P_n(x) = -n(n+2)xP_n(x) \quad (17)$$

there are solutions which are analytic at both ends of the interval $[0,1]$ and whose power series terminate. These functions are related to Jacobi polynomials, and one can work out all their properties and expansions using standard analysis methods.

The reader may object that we are not implementing conjugate gradient at all, and so we may learn nothing about the convergence of the method. While it is true that we are avoiding the explicit use of conjugate gradient, we have nonetheless determined a sequence of polynomials which have all the properties of the conjugate gradient sequence. Since there can only be one such, we are using the same expansion vectors as would have been given by a direct use of conjugate gradient. That is, the convergence of the Sturm-Liouville expansion is exactly the convergence of the conjugate gradient expansion, because they are the same expansion.

A convenient normalization for these polynomials is

$$P_n(x) = \sum_{m=1}^{n} \frac{(-1)^{m-1}(n-1)!}{(m-1)!(n-m)!} \frac{3!(n+m+1)!}{(n+2)!(m+2)!} x^m \quad (18)$$

Note that the sequence starts at $n = 1$, not $n = 0$. The constant function is not in the sequence. This is what makes the problem of expanding the exact solution of the minimization problem, the constant function, other than trivial. It is straightforward to derive the integrals for single functions and for a product of two:

$$\int_0^1 x P_n(x)\,dx = \frac{12}{[n(n+1)(n+2)]^2}$$
$$\int_0^1 x P_m(x) P_n(x)\,dx = \frac{18}{n^2(n+1)^3(n+2)^2}\delta_{mn}$$
(19)

These integrals are used in evaluating the conjugate gradient expansion coefficients and examining the convergence in the norm of the partial sums. Explicitly, the successive approximations $f_n(x)$ that converge to $f(x) = 1$ are given by

$$f_n(x) = \frac{2}{3}\sum_{m=1}^{n}(n+1)P_m(x)$$
$$= 1 - 2\sum_{m=0}^{n}\frac{(n!)^2(-x)^m(1-x)^{n-m}}{m!(m+2)!((n-m)!)^2}$$
(20)

## 6.2 Convergence Properties

The convergence of the partial sums in the norm is only a power of the number of terms,

$$\int_0^1 x \, (f_n(x) - 1)^2 = \frac{2}{[(n+1)(n+2)]^2} \quad (21)$$

The number of terms needed to successively reduce the error by a fixed finite factor grows like the $-1/4$ power of the residual error, so that the relaxation time of the sequence is infinite.

The pointwise convergence is both slow and non-uniform. At fixed $x$ in the interior of $[0,1]$ one has

$$f_n(x) - 1 \sim \frac{\cos(n\theta + c)}{n^{-5/2}} \quad (\cos\theta \equiv 1 - 2x) \quad (22)$$

At the upper end of the interval the convergence is different:

$$f_n(1) = \frac{2(-1)^n}{(n+1)(n+2)} \quad (23)$$

This is a marginally slower rate of convergence. At the lower end of the interval, pointwise convergence actually fails.

## 6.3 The Gibbs Phenomenon

No matter how many terms $n$ are kept in the expansion, there remains a region in which the difference between the partial sum and the exact result does not approach 0. The region shrinks as $n$ grows, of course. Specifically, for small values of $x \propto 1/n^2$, one derives

$$\lim_{n \to \infty} f_n\left(\frac{a}{n^2}\right) = 1 - \frac{2}{a} J_2(2\sqrt{a}) \quad (24)$$

where, $J_2$ denotes the ordinary (oscillatory) Bessel function. The maximum value of this expression is about $1.059$, which is attained at $a \approx 1.8$

This is an example of the Gibbs phenomenon. The series overshoots and oscillates about true value. Even though this behavior is restricted to an ever smaller range, $x \propto 1/n^2$, it saturates the convergence in the norm. Explicitly, one may show that

$$\lim_{b \to \infty} \frac{1}{n^4} \int_0^b a \, da \lim_{n \to \infty} \left(f_n\left(\frac{a}{n^2}\right) - 1\right)^2 = \frac{2}{n^4} \quad (25)$$

which is the leading asymptotic behavior of the full deviation in the norm from $f(x) = 1$.

## 6.4 Role of the Small Eigenvalues

The appearance of the Gibbs phenomenon, and the turgid convergence in general, are a result of the multiplicity of small eigenvalues. It is instructive to consider what would be the convergence of conjugate gradient if one chose the matrix $M_{ij}$ not to have any small eigenvalues. In the $N \to \infty$ limit, the effect would be to replace the polynomials $P_n(x)$ with polynomials $K_n(x)$ whose orthogonality relation involves a weight function that does not take arbitrarily small values. As an example, let us consider polynomials orthogonal with respect to

$$w(x) = (x + 1) \quad (26)$$

These polynomials are not related to any of the classical polynomials, and so we have no short cuts to assist the evaluation of the expansion coefficients or the convergence properties of the partial sums. Nonetheless, it is a simple computer algebra exercise to compute the first dozen or two terms exactly. The result shows numerically that the successive approximations to $f(x) = 1$ converge exponentially with $n$. The following table shows the contrast between the singular expansion in the $P_n$ polynomials and the smooth expansion in the $K_n$ polynomials

Table 1: Convergence of $\int_0^1 w(x)\,(f_n(x) - 1)^2\,dx$

| Order | P - Expansion | K - Expansion |
|-------|---------------|---------------|
| 0     | 1.            | 1.            |
| 1     | .2592593      | 0.03670782    |
| 2     | .1203704      | 0.001182794   |
| 3     | .07           | 3.597242e-5   |
| 4     | .04592593     | 1.077474e-6   |
| 5     | .03250189     | 3.206815e-8   |
| 6     | .02423469     | 9.511775e-10  |
| 7     | .01877572     | 2.815558e-11  |
| 8     | .01497942     | 8.32333e-13   |
| 9     | .0122314      | 2.458333e-14  |
| 10    | .01017753     | 7.256163e-16  |

## 7 Conclusions

The principal lesson of this analysis is qualitative. It is that the slowness of the gauge fixing process to the lattice Landau gauge for large lattices is not likely to be overcome. Even global maximization methods, and even neglecting their limitations imposed by round-off errors, do not seem to help. To a lesser extent, the same observation could be made about the lattice Coulomb gauge; lesser simply because the number of sites to be fixed is smaller by a factor of the number of sites in the time direction.

One way to try to deal with difficulty of gauge fixing on very large lattices would be via the use of perfect, or at least improved actions. That is, one may try to use lattices with many fewer sites, for the same level of precision. This will not be automatic to implement however. Just as additional operators in the action are needed to eliminate the leading finite lattice spacing errors, similar terms will be required in the gauge condition, to improve the convergence of the lattice Landau gauge to the continuum version.

Another necessary observation about extending the improvement program to gauge-variant quantities is that all the evidence for the viability of that program comes from the study of gauge-invariant quantities. It will be necessary to check carefully the extension of improved actions and operators to the unphysical sectors of gauge theory. A most non-trivial test of those ideas will be whether the improvement of the action and operators with gauge-invariant improvement terms suffices, or whether non-gauge-invariant improvement terms will need to be introduced.